\documentclass[11pt]{article}
\usepackage{epsfig}
\usepackage{amssymb}
\usepackage{amsmath}

\newcommand{\ee}{\end{equation}}
\newcommand{\eea}{\end{eqnarray}}
\newcommand{\be}{\begin{equation}}
\newcommand{\bea}{\begin{eqnarray}}

\begin{document}

\title{\LARGE \bf Self-Gravitating Spherical Solutions of the non-minimally coupled non-Abelian Higgs Model}
  \author{
  \large  Y. Brihaye$^a$ $\:${\em and}$\:$
  Y. Verbin$^b$ \thanks{Electronic addresses: yves.brihaye@umons.ac.be; verbin@openu.ac.il } }
 \date{ }
   \maketitle
    \centerline{$^a$ \em Physique Th\'eorique et Math\'ematiques, Universit\'e de Mons,}
   \centerline{\em Place du Parc, B-7000  Mons, Belgique}
     \vskip 0.4cm
   \centerline{$^b$ \em Department of Natural Sciences, The Open University
   of Israel,}
   \centerline{\em Raanana 43107, Israel}
%   \vskip 0.5cm

%%%%%%\begin{titlepage}
\maketitle
\thispagestyle{empty}
\begin{abstract}
Motivated by the Higgs Inflation scenario, we study static spherically-symmetric solutions of the non-Abelian Higgs model  coupled non-minimally to Gravity. We find
solutions for the self-gravitating sphaleron as well as monopole-like solutions and study the impact of the non-minimal coupling on their properties. Finally we discuss shortly the possibility that these solutions interact gravitationally with star-like objects like boson stars.
\end{abstract}

\maketitle
%\medskip \medskip


\section{Introduction}\label{Introduction}
\setcounter{equation}{0}

The Higgs inflation scenario \cite{BezrukovEtAl2007,BezrukovEtAl2008,BezrukovEtAl2010} is probably the most ``parsimonious'' version of inflation in the sense that it does not require any additional fields beyond those of the Standard Model (SM) of particle physics. This is achieved by letting the standard model Brout-Englert-Higgs field to play the role of the inflaton and further assuming a non-minimal coupling of this bosonic  field to gravity, represented by an additional  $-\xi R |\Phi|^2$ term in the Lagrangian. Another facet of the parsimoniousness of this model is the fact that it fills solely the ``zero parameter model'' slot as classified by Encyclop{\ae}dia Inflationaris \cite{EncyclopInfl2014}.

This model passes all the present observational cosmological tests. However, there are additional tests it should pass. First and foremost is the impact of the non-minimal coupling on smaller scale structures from star-like objects to solitons. One such study \cite{Fuzfa:2013yba,Schlogel:2014jea} was carried out recently with the (quite expected) conclusion that the additional coupling has negligible effects on stellar
structures or more specifically, on stars with global U(1) scalar ``hair''. This result holds even when the non minimal coupling parameter $-\xi$ is as large as $10^4$ as required by the Higgs inflation models.

The impact on self-gravitating solitonic solutions like magnetic monopoles, sphalerons and others has not been studied extensively so far, and we would like to report here on some results about the effect of the non-minimal coupling on self-gravitating magnetic monopoles and sphalerons. Indeed, the original motivation of suggesting the Higgs inflation scenario was to obtain inflation without the need of any beyond SM fields. However, this possibility is still open and therefore we will not limit ourselves to the domain of the weak scale of the order of $10^2$GeV, but allow ourselves to consider energy scales up to the GUT scale of around $10^{15}$GeV.

Self-gravitating magnetic monopoles and sphalerons appear naturally \cite{GaltsovVolkov} in the non-Abelian Higgs model whose field content is:
\begin {itemize}
\item {Scalar field $\Phi$ which transforms according to some representation of the gauge group generated by $n$ generators $T^a$ , $a=1,...,n$.}
\item {Lie algebra-valued gauge potential $A_\mu dx^\mu=T^a A^a_\mu dx^\mu$. $A^a_\mu dx^\mu$ may be viewed as the $n$ components of the gauge potential.}
\end {itemize}
%%% Change conventions%%%
The Lagrangian which we will use here is
\begin{equation}
{\cal L}= (D_\mu \Phi)^\dagger(D^\mu\Phi) -\frac{\lambda}{4}\left(|\Phi|^2-\frac{v^2}{2}\right)^2
-\frac{1}{4}F^a_{\mu\nu}F^{a \mu\nu} -\xi R|\Phi|^2 + \frac{1}{2\kappa}R +
{\cal L}_{add} \label{TotLag},
\end{equation}
where $\kappa = 8 \pi G$, $D_\mu = \nabla_\mu - ieA^a_\mu T^a$ and $F^a_{\mu\nu}$ are the $n$ components of the Lie algebra-valued field strength $F^a_{\mu\nu} T^a$.  Our sign convention is such that the conformal coupling corresponds to $\xi=1/6$ - see the simplifications in Eq (\ref{Rintermsofrest}) below. Observational data require $\xi$ to be of the order of $10^4$. ${\cal L}_{add}$ is an additional Lagrangian for a possible additional matter which does not interact with the Higgs system. We will consider briefly this possibility in sec. \ref{S-BS}.

The resulting field equations of the Yang-Mills-Higgs (YMH) system are
\begin{eqnarray}
D_\mu D^\mu \Phi + \frac{\lambda}{2}\left(|\Phi|^2-\frac{v^2}{2}\right) \Phi + \xi R \Phi &=& 0 \label{FieldEqsScalar}\\
D_\mu F^{a \mu\nu} =-ie[\Phi^\dagger T^a (D^\nu \Phi)- (D^\nu \Phi)^\dagger T^a \Phi ]&=& J^{a\nu }
\label{FieldEqsVector}.
\end{eqnarray}

The gravitational field equations are
\begin{equation}
(1-2\xi \kappa |\Phi|^2){G}_{\mu\nu} = -\kappa \left ({ T}_{\mu\nu}^{(H)}+2\xi
\left (g_{\mu \nu} \nabla ^{\lambda }\nabla_
{\lambda } |\Phi|^2 - \nabla _{\mu }\nabla_{\nu } |\Phi|^2 \right)\right)
\label{confTmn}
\end{equation}
${T}_{\mu\nu}^{(H)}$ being the ordinary (``minimal'') energy-momentum tensor of the
YMH model and ${G}_{\mu \nu}$ is the Einstein tensor. An additional energy-momentum tensor ${T}_{\mu\nu}^{(add)}$ may appear in the RHS if additional matter is introduced.

We write here also a useful relation which will be used to simplify eq. (\ref{FieldEqsScalar}) by expressing $R$ in terms of the other degrees of freedom:
\begin{equation}
R=\frac{2(6\xi-1)(D_\mu \Phi)^\dagger(D^\mu\Phi)-\lambda  (|\Phi|^2-v^2 /2)[(6\xi-1)|\Phi|^2+v^2 /2]+T^{(add)}}{1/\kappa + 2\xi(6\xi-1)|\Phi|^2}
\label{Rintermsofrest}
\end{equation}

Now we concentrate in spherical symmetry and choose the following parametrization for the line element:
\begin{equation}
ds^2= A^2 (r) N(r)dt^2 - dr^2/N (r) - r^2 ( d\theta^2+\sin^2 \theta d\varphi^2)
\label{lineelsph}.
\end{equation}
The non-vanishing components of the Einstein tensor are then
\begin{eqnarray}
G^0_0 = \frac{N'}{r}+\frac{N-1}{r^2} \,\,\,\,\,\,; \,\,\,\,\,\,
G^1_1 = \frac{N'}{r}+\frac{N-1}{r^2}+\frac{2N}{r}\frac{A'}{A} \,\,\,\,\,\,; \,\,\,\,\,\,\\ \nonumber
G^\theta_\theta = G^\varphi_\varphi =
\frac{N''}{2}+\frac{N'}{r}+\frac{N A''}{A}+ \left(\frac{3 N'}{2}+\frac{N}{r}\right)\frac{A'}{A}
\label{SphEinst}.
\end{eqnarray}

For the gauge group we take the simplest SU(2) with the spherically-symmetric gauge field
\begin{equation}
A=A_\mu dx^\mu = \frac{f(r)-1}{er^2} \epsilon_{aij} T_a x^i dx^j = \frac{f(r)-1}{e}(T_\varphi d\theta - T_\theta \sin \theta d\varphi)
\label{GaugeFsph}
\end{equation}
where $a,i,j$ take values of ${1,2,3}$ and $x^i$ are the 3 cartesian coordinates. $T_a$ are the 3 generators of SU(2) in the appropriate representation while  ($T_r, T_\theta, T_\varphi$) are the ``spherical'' ones, namely just products with the standard spherical unit vectors in 3-space, $\textbf{e}_r$, $\textbf{e}_\theta$ and $\textbf{e}_\varphi$. The resulting field strength has the following non-vanishing components which are written very simply as
\begin{eqnarray}
F_{r\theta} = -F_{\theta r} =\frac{f'}{e} T_\varphi\,\,\,\,\,\,; \,\,\,
F_{\varphi r} =-F_{r\varphi} = \frac{f'}{e} T_\theta \sin \theta\,\,\,\,\,\,; \,\,\,
F_{\theta\varphi}=-F_{\varphi\theta}=\frac{f^2-1}{e} T_r \sin \theta\
\label{SphGField}.
\end{eqnarray}

As for the scalar field, there are two well-known spherically-symmetric possibilities: the sphaleron which is an isospin 1/2 representation (i.e. $T_a=\sigma_a /2$) and the monopole which corresponds to isospin 1 (i.e. $(T_a)_{bc}=-i\epsilon_{abc}$) ``hedgehog'' configuration:
\begin{eqnarray} \label{Sphaleron+Monopole}
\Phi_{sphal} = \frac{i v}{\sqrt{2}} K(r) \sigma_r \left(\begin{array}{c}
 0 \\
 1 \\
\end{array}
\right)  = \frac{i v}{\sqrt{2}} K(r) \left(\begin{array}{c}
 \sin \theta e^{-i\varphi} \\
 -\cos \theta \\
\end{array}
\right) \\ \nonumber
 \Phi_{mon} = \frac{v}{\sqrt{2}} H(r)\textbf{e}_r = \frac{v}{\sqrt{2}} H(r) \left(\begin{array}{c}
\sin\theta \cos\varphi \\
\sin\theta \sin\varphi\\
 \cos\theta \\
\end{array}
\right).
\end{eqnarray}

There exists a previous study of sphalerons with non-minimal coupling to gravity by van der Bij \& Radu  \cite{vanderBij:2000cu}. It is however limited to the demonstration of the existence of these solutions and drawing some qualitative characteristics. Here we expand the discussion and chart more extensively the space of solutions in view of the Higgs inflation scenario. We were unable to find in the literature an analogous inquiry into the problem of the magnetic monopole - not even in the simplest case of conformal coupling ($\xi=1/6$).  In this respect the results we present here are even more innovative.
The closest are the three papers by Wali and collaborators \cite{Wali+Balakrishna,Wali+Nguyen,WaliEtAl2002} that omit from the action the Einstein-Hilbert term  altogether.

\section{Field Equations for the Sphaleron}
\label{FEqs-Sphaleron}
\setcounter{equation}{0}

Substituting the above for isospin 1/2 into the field equations and performing some simplifications, yield the following system for the 4 unknown functions $A(r)$, $N(r)$, $K(r)$ and $f(r)$. For the YMH system we obtain:
\begin{eqnarray}\label{HiggsEqs}
 \frac{(r^2AN K')'}{r^2 A}-\frac{(f+1)^2}{2r^2} K - \frac{\lambda}{4} v^2 (K^2-1)K +
 \, \hspace {6.0cm} \\ \nonumber
 \xi \frac{(6\xi-1)[N(K')^2+\frac{(f+1)^2 K^2}{2r^2}]+\frac{\lambda}{4} v^2 (K^2-1)[(6\xi-1)K^2+1]}{1/\kappa v^2 + \xi(6\xi-1)K^2}K=0
\end{eqnarray}
and
\begin{equation}
\frac{(ANf')'}{A}-\frac{(ev)^2}{4} K^2 (f+1) - \frac{(f^2-1)f}{r^2}=0
\label{GaugeEqs}.
\end{equation}
The (00) and (11) gravitational field equations turn out to be first order:
\begin{eqnarray}
\frac{N'}{r}+\frac{N-1}{r^2}+\frac{\kappa\left(\varepsilon_0+\varepsilon_1+u_0 +u_1+u_2-\xi(\tau_0+4\varepsilon_0-\tau_1)\right)}{1-\xi\kappa v^2 K^2}=0
\label{EinsEq00}
\end{eqnarray}
\begin{eqnarray}
\frac{N'}{r}+\frac{N-1}{r^2}+\frac{2N}{r}\frac{A'}{A}+\frac{\kappa\left(-\varepsilon_0-\varepsilon_1+u_0 +u_1+u_2-\xi\tau_2\right)}{1-\xi\kappa v^2 K^2}=0
\label{EinsEq11}
\end{eqnarray}
where
\begin{eqnarray}
\varepsilon_0=\frac{v^2N(K')^2}{2} \,\,\,\,\,\,&;& \,\,\, \varepsilon_1=\frac{N(f')^2}{e^2 r^2} \\ \nonumber
u_0=\frac{\lambda v^4}{16} (K^2-1)^2  \,\,\,\,\,\,&;& \,\,\, u_1=\frac{(f^2-1)^2}{2e^2 r^4} \,\,\,\,\,\,; \,\,\, u_2=\frac{v^2(f+1)^2 K^2}{4r^2}\\ \nonumber
\tau_0=\frac{2v^2 K(r^2AN K')'}{r^2 A} \,\,\,\,\,\,&;& \,\,\,\tau_1=2v^2NKK'\left(\frac{A'}{A}+\frac{N'}{2N}\right)
\,\,\,\,\,\,; \,\,\,\tau_2=\tau_1+\frac{4v^2 NKK'}{r}
\end{eqnarray}
Actually, Eq (\ref{EinsEq00}) is second order in $K$ because of the $\tau_0$ term, but it can be easily converted to first order by using the field equation (\ref{HiggsEqs}) for $K$ as we will do below.

The first physical quantity of interest is the mass. The shortest way to the mass is to write
$N(r)=1-2{\cal M}(r)/r$ where ${\cal M}(r)$ is the cumulative mass function such that the mass $M$ is extracted from the asymptotic behavior of $N(r)$: $GM={\cal M}(\infty)$. Alternatively an integral expression may be obtained using Eq (\ref{EinsEq00}):
\begin{eqnarray}
M=4\pi \int_0^{\infty} dr \hspace {0.15cm} r^2 \hspace {0.1cm} \frac{\varepsilon_0+\varepsilon_1+u_0 +u_1+u_2-\xi(\tau_0+4\varepsilon_0-\tau_1)}{1-\xi\kappa v^2 K^2}
\label{MassSphaleron}
\end{eqnarray}
Notice the presence of the non-minimal coupling in this expression.

\vspace {0.25cm}
%\clearpage
\textbf{Field Equations in Dimensionless Form}: By defining $x=evr/\sqrt{2}$, $\alpha^2=\kappa v^2 /4$ and $\beta=\frac{\lambda}{e^2}$ we get the following system in dimensionless form with the 3 free parameters $\alpha$, $\beta$ and $\xi$. The YMH equations:
\begin{eqnarray}\label{HiggsEqsDless}
 \frac{(x^2AN K')'}{x^2 A}-\frac{(f+1)^2}{2x^2} K - \frac{\beta}{2} (K^2-1)K +
  \hspace {6.0cm} \\  \nonumber
 4\xi \alpha^2 \frac{(6\xi-1)[N(K')^2+\frac{(f+1)^2 K^2}{2x^2}]+\frac{\beta}{2} (K^2-1)[(6\xi-1)K^2+1]}{1 + 4\xi \alpha^2 (6\xi-1)K^2}K=0
\end{eqnarray}
\begin{equation}
\frac{(ANf')'}{A}-\frac{1}{2} K^2 (f+1) - \frac{(f^2-1)f}{x^2}=0
\label{GaugeEqsDless},
\end{equation}
and the (00) and (11) gravitational field equations:
\begin{eqnarray}
\frac{N'}{x}+\frac{N-1}{x^2}+\frac{2\alpha^2}{1-4\xi\alpha^2 K^2}\left[N(K')^2+\frac{N(f')^2}{x^2}+\frac{\beta}{4} (K^2-1)^2  +\frac{(f^2-1)^2}{2x^4} \nonumber \right. \\ \left.+\frac{(f+1)^2 K^2}{2x^2}-4\xi \left(\frac{K(x^2AN K')'}{x^2 A}+N(K')^2-NKK'\left(\frac{A'}{A}+\frac{N'}{2N}\right)\right)\right]=0
\label{EinsEq00Dless}
\end{eqnarray}
\begin{eqnarray}
\frac{N'}{x}+\frac{N-1}{x^2}+\frac{2N}{x}\frac{A'}{A}+\frac{2\alpha^2}{1-4\xi\alpha^2 K^2}\left[-N(K')^2-\frac{N(f')^2}{x^2}+\frac{\beta}{4} (K^2-1)^2 \nonumber \right. \\ \left. +\frac{(f^2-1)^2}{2 x^4}+\frac{(f+1)^2 K^2}{2x^2}-4\xi NKK' \left(\frac{A'}{A}+\frac{N'}{2N}+\frac{2}{x}\right)\right]=0
\label{EinsEq11Dless}
\end{eqnarray}

As mentioned already, Eq (\ref{EinsEq00Dless}) can be easily converted to first order (in $K$ too) by using the field equation (\ref{HiggsEqsDless}) for $K$ and writing:
\begin{eqnarray}
\frac{K(x^2AN K')'}{x^2 A}= \frac{\frac{(f+1)^2}{2 x^2}-\frac{\beta}{2} \left( 4\xi \alpha ^2 -1\right) \left(K^2-1\right)  -  4\xi  \alpha ^2 (6 \xi -1) N
   \left(K'\right)^2}{1+  4\xi \alpha^2 (6 \xi -1) K^2 } K^2
\label{K''substit}
\end{eqnarray}

Finally we note that the dimensionless mass parameter $\mu$ which is the coefficient of the $2/x$ term in the asymptotic expansion of $N(x)$ is just ${\cal M}(\infty)ev/\sqrt{2} = GMev/\sqrt{2}$. Thus $\bar{M}=\alpha^2\mu \sqrt{2}$ gives the mass in units of $M_W/ \alpha_w$ where $M_W = ev/2$ and $\alpha_w = e^2/4\pi$.

\section{Numerical Results for the Sphaleron}
\label{Numerical-Sphaleron}
\setcounter{equation}{0}
Since there are, to our knowledge, no closed form solutions for the system under consideration,
we have solved the equations numerically by using the solver COLSYS \cite{colsys}.
\subsection{Sphalerons with minimal coupling}
We first reexamine the solutions minimally coupled to gravity. It is known
for some time \cite {Greene:1992fw} that several  families of solutions exist on finite intervals
of the effective coupling constant $\alpha$. These families are labeled by the number
of zeros, say $k$ of the function $f$.  The case $k=0$ corresponds to the vacuum solution $f = -1$, $K = 1$.
The boundary conditions of the matter fields are different according to the parity of the integer $k$.
For $k=1$ and $k=2$, we have respectively
\be
            f(0) = 1 \ \ , \ \ K(0) = 0 \ \ , \ \ f(\infty) = -1 \ \ , \ \ K(\infty) = 1 \ \ ,
\ee
\be
            f(0) = -1 \ \ , \ \ K'(0) = 0 \ \ , \ \ f(\infty) = -1 \ \ , \ \ K(\infty) = 1 \ \ .
\ee
The $k=1$ family corresponds to the gravitating version of the
Klinkhamer-Manton sphaleron \cite{Klinkhamer:1984di}. The family in fact consists of two branches
which exist for $\alpha \in [0, \alpha_m(1)]$ where $\alpha_m(1)$ depends on $\beta$.
Fixing a value for $\alpha$ in this interval, two  solutions exist which
 are distinguished by their masses and other characteristics as shown in Fig.\ref{data_sphaleron_0}(a) for $\beta = 0.5$. Profiles of a typical solution are shown (as the dashed curves) in Fig.\ref{inflation_0}. The two branches coincide for $\alpha \to \alpha_m(1)$.
 In the limit $\alpha \to 0$, the branch with the lowest mass,
  call it $br_l$ approaches the  sphaleron in flat space, the branch with the higher mass approaches the first
 solution of the Bartnik-MacKinnon family \cite{BMK1988} (at least with an appropriate rescaling of the radial variable).

 The solutions with the lowest mass are  the most physically interesting ones since they
 represent the energy barrier between topologically inequivalent vacua of the underlying gauge field theory. In the following, we will pay a special attention to the effect of the non-minimal coupling  on the energy barrier.

 It should be pointed out that the masses of the solutions of the two branches decrease while
 the parameter $\alpha$ increases. Setting for definiteness $\beta = 0.5$, we find for the 'sphaleron'
  branch $br_l$ the values
 $\bar{M} \approx 3.63$ and $\bar{M} \approx 3.03$  (in units of $M_W/ \alpha_w$) respectively for $\alpha = 0$ and $\alpha = \alpha_m(1) \approx 0.43$.

Throughout our numerical analysis, we set $\beta = 0.5$ which, in our scale, is close to the experimental value
of the Higgs boson.   We believe that this value reflects the qualitative pattern of the solutions for
the low values of $\beta$. We do not include in our present study larger values of
 $\beta$ which allow for other types of solutions, for instance the bisphaleron solutions \cite{Kunz:1988sx,Yaffe:1989ms}.

\begin{figure}[h]
\begin{center}
%%%\subfigure[Profile]
{\includegraphics[width=12cm]{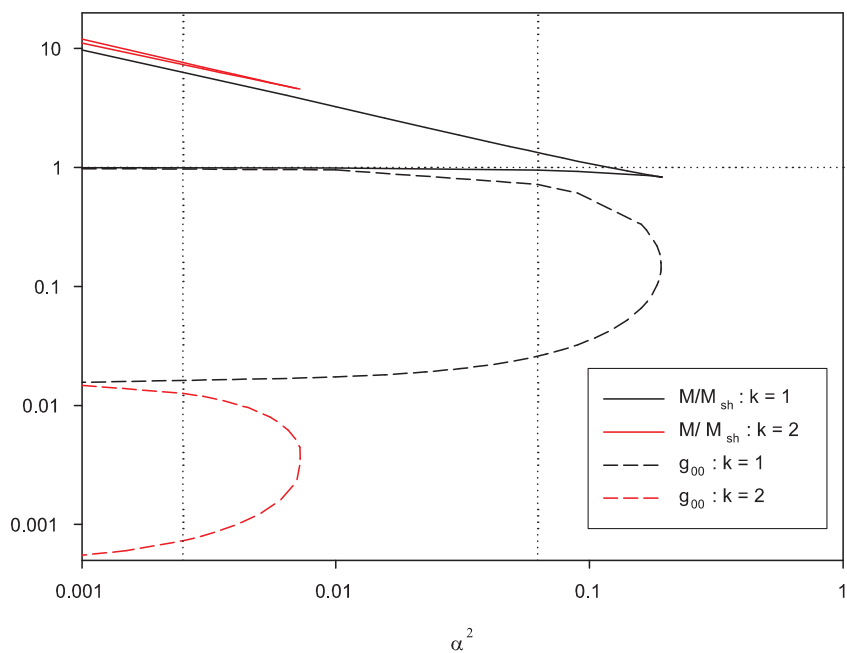}}
%%%\subfigure[Particle number, Mass and mean radius]
%%%{\label{hyper_r}\includegraphics[width=8cm]{hyper.eps}}
\end{center}
\caption{Mass and $g_{00}(0)$ of the minimally-coupled ($\xi=0$) $k=1$ and $k=2$ families as function of $\alpha^2$ for $\beta = 0.5$.
\label{comparaison_k_1_2}
}
\end{figure}

  Now we discuss shortly  the family corresponding to $k=2$. This family also occurs in the form of two branches which exist for $\alpha \in ]0, \alpha_m(2)]$. In the limit $\alpha \to 0$, the solutions in one of these branches
 corresponding to the lowest energy,
 approach the second Bartnik-MacKinnon solution (again up to a rescaling of the radial variable).
 With the generic value $\beta = 0.5$,
 we find $\alpha_m(2) \approx 0.08$.
A comparison between the $k=1$ and $k=2$ branches is presented in Fig.\ref{comparaison_k_1_2}. Profiles of a typical solution are shown (as the dashed curves) in Fig.\ref{inflation_1}.

%%% $\alpha \in [0, \alpha_m(k)]$.
\subsection{Sphaleron with non minimal coupling}
\begin{figure}[h]
\begin{center}
%%%\subfigure[Profile]
{\includegraphics[width=12cm]{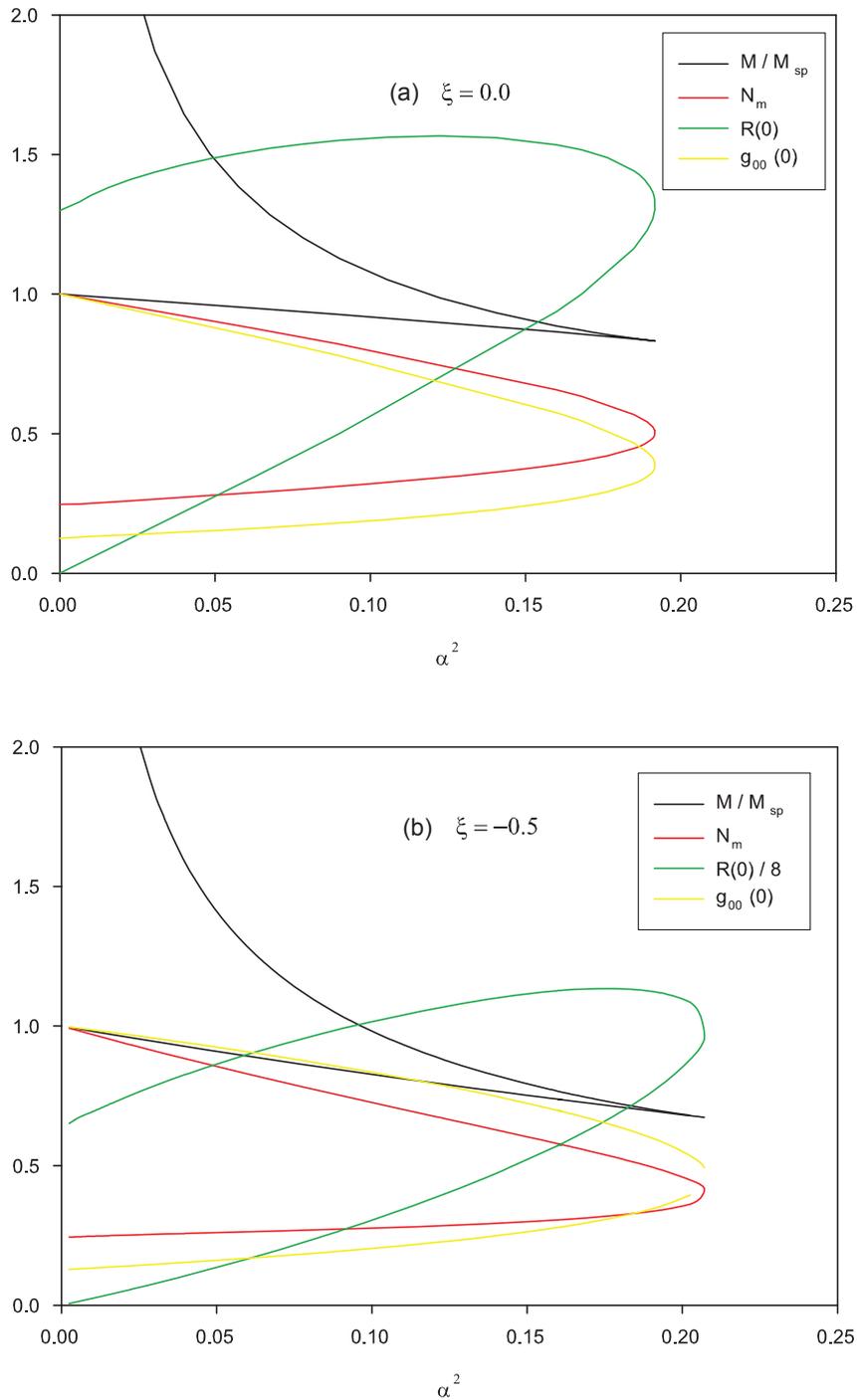}}
%%%\subfigure[Particle number, Mass and mean radius]
%%%{\label{hyper_r}\includegraphics[width=8cm]{hyper.eps}}
\end{center}
\caption{Several quantities characterizing the sphaleron as functions of the parameter $\alpha$ for $\beta=0.5$ and: (a) $\xi = 0$; (b)  $\xi = - 0.5$ . The quantities are the mass $M$ in units of the flat space sphaleron mass designated $M_{sp}$, $N_m$, the minimal value of $N(x)$, $R(0)$ and $g_{00}(0)$.
\label{data_sphaleron_0}
}
\end{figure}

\begin{figure}[h]
\begin{center}
%%%\subfigure[Profile]
{\label{data_sphaleron_l}\includegraphics[width=12cm]{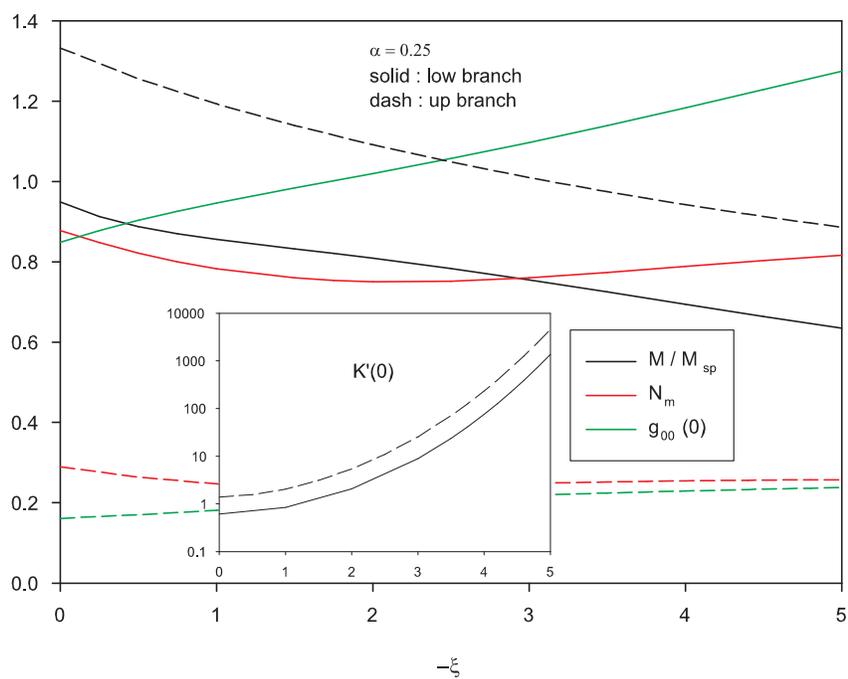}}
%%%\subfigure[Particle number, Mass and mean radius]
%%%{\label{hyper_r}\includegraphics[width=8cm]{hyper.eps}}
\end{center}
\caption{Several quantities characterizing the sphaleron as functions
of the parameter $\xi$ for $\alpha = 0.25$ and $\beta=0.5$: Mass in units of the flat space sphaleron mass, $g_{00}(0)$, $N_m$ and $K'(0)$.
\label{data_sphaleron_1}
}
\end{figure}

\begin{figure}[h]
\begin{center}
%%%\subfigure[Profile]
{\includegraphics[width=12cm]{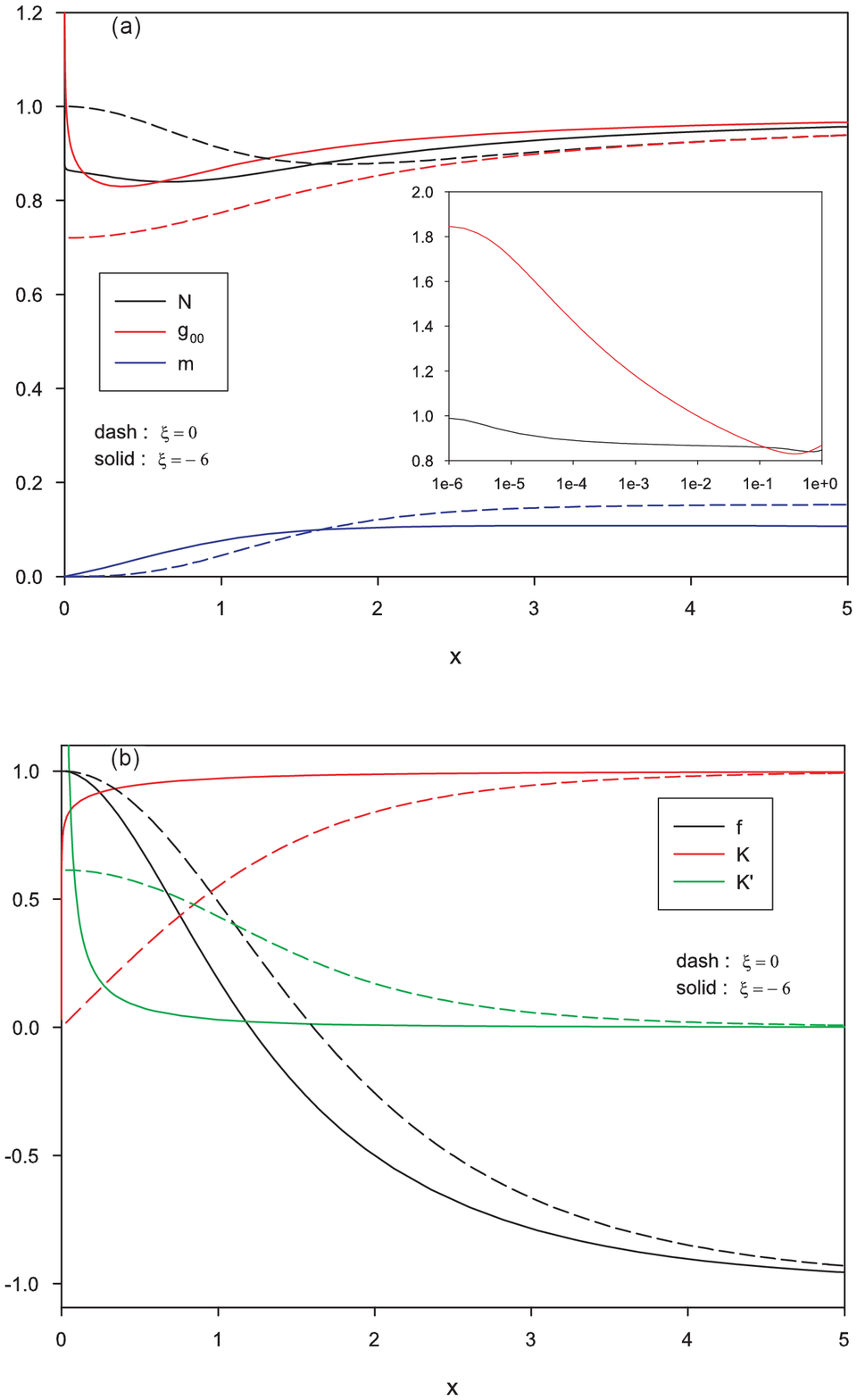}}
%%%\subfigure[Particle number, Mass and mean radius]
%%%{\label{hyper_r}\includegraphics[width=8cm]{hyper.eps}}
\end{center}
\caption{Profiles of a typical $k=1$ sphaleron for minimal ($\xi = 0$) and non-minimal coupling with $\xi = - 6$. In both cases $\alpha = 0.25$ and $\beta=0.5$. (a) metric functions; (b) YMH fields.
The insert in (a) magnifies the region near the origin and demonstrates the smooth behavior of the metric components for $\xi < 0$ which is not obvious from the larger scale curves. The function $m(x)$ is defined by $N(x)=1-2m(x)/x$.
\label{inflation_0}
}
\end{figure}

For the $k=1$ family,
the pattern discussed above seems to be preserved for small values of $|\xi|$. This is illustrated in Fig.\ref{data_sphaleron_0}(b)
for  $\xi = - 0.5$ . The purpose of this plot is to
demonstrate the effect of a small value of the non minimal constant $\xi$ on the sphaleron. It reveals in particular
that even a small value of the parameter $\xi$ leads to a significant increase of the Ricci scalar in the center of the sphaleron. From the following expression for $R(0)$
\be
R(0)=2(ev)^2\alpha^2\left[ (1-6\xi)(K'(0))^2 + \beta/2 \right]
\ee
it is obvious that it is not the explicit appearance of $\xi$ that is responsible for the strong increase of $R(0)$, but rather the way $K'(0)$ depends on $\xi$. This is obvious in Fig.\ref{data_sphaleron_1}. The large curvature around the origin is also reflected in Fig.\ref{inflation_0}  by the strong increase in $N(r)$, $g_{00}(r)$ and $K'(r)$ towards $r\rightarrow 0$.

It is also natural to study the dependance of the  gravitating sphaleron on
 $\xi$ for a fixed value of the parameter $\alpha$.
This analysis reveals one feature of the pattern:  when the parameter $\xi$ is decreased (recall: $\xi<0$, so $-\xi$ increases),
the scalar field becomes more and more
concentrated around the origin while the gauge field remain roughly $\xi$ independent. This is also in line with the absence of $\xi$ from the field equation (\ref{GaugeEqsDless}) for $f(x)$.
In particular  the value $K'(0)$ increases considerably when $\xi$ is decreased;
correspondingly the value $R(0)$ also becomes very large, rendering
the numerical integration quite difficult.
for  $0 < \xi < 1$ we find solutions which smoothly deform the $\xi = 0$ solutions,
  the pattern in the two branches seems to be the same. However, since the emphasis
  of the paper is to investigate the Higgs inflation region $\xi<0$, the full study of the case $\xi > 0$ was not addressed here.

We present in Fig.\ref{data_sphaleron_1}
 a few parameters characterizing the sphaleron non minimally coupled
to gravity and with a generic value of $\alpha$ (for instance $\alpha = 0.25$); the solid (resp. dashed)
curves refer to data corresponding to the lower (resp. higher) solutions . The insert illustrates the increase of $K'(0)$ (note the logarithmic scale).
The profiles of a typical solution (shown in Fig.\ref{inflation_0}) reveal  that for $\xi <0$ the metric
$g_{00}(0)$ gets larger than 1 and that both the potential $g_{00}(r)$ and the function $N(r)$
exhibit a local minimum before reaching their asymptotic value $g_{00}(r \to \infty)=N(r \to \infty)=1$.

Completing Fig.\ref{inflation_0} by $\xi = 0$ plots,
the difference between the minimally and non-minimally coupled sphalerons can be appreciated.

%%% An illustration of this phenomenon is presented on Fig. \ref{data_sphaleron_1} where $\alpha = 0.25$ is set.
%%% The numerical integration becomes hard for $\xi < -6$ but it is likely that for $\xi << -5$, the
%%% scalar function $K(r)$ approaches the function $K(r)=1$ for $r \in ]0,\infty]$ with a jump at the origin.
%%% Figure \ref{inflation_0} shows an illustration of this claim

%%%****\emph{NEW ADDITION 14 OCT}**** \\

The discussion above, just leads to a reasonable understanding of the effect of the different
gravitational couplings $\alpha, \xi$. According to \cite{BezrukovEtAl2007} the physically relevant
domain occurs for  $\xi \sim 10^{4}$ quite independently of $\alpha$ as long as it is small enough, i.e. from $\alpha \sim 10^{-17}$ corresponding to the electroweak scale up to the GUT value of $\alpha \sim 10^{-4}$.

It is of course very much expected that the non-minimal coupling will have a negligible effect on the weak scale sphalerons and monopoles. So in order to understand the role of the non-minimal coupling, higher energy scale should be assumed.

%\underline{DELETE BETWEEN *** ***:} *** It turns out unrealistic to solve the system of equations with %two such different scales but we report here a set of numerical data which suggest what could be % expected. ***

\begin{figure}[h]
\begin{center}
%%%\subfigure[Profile]
{\includegraphics[width=12cm]{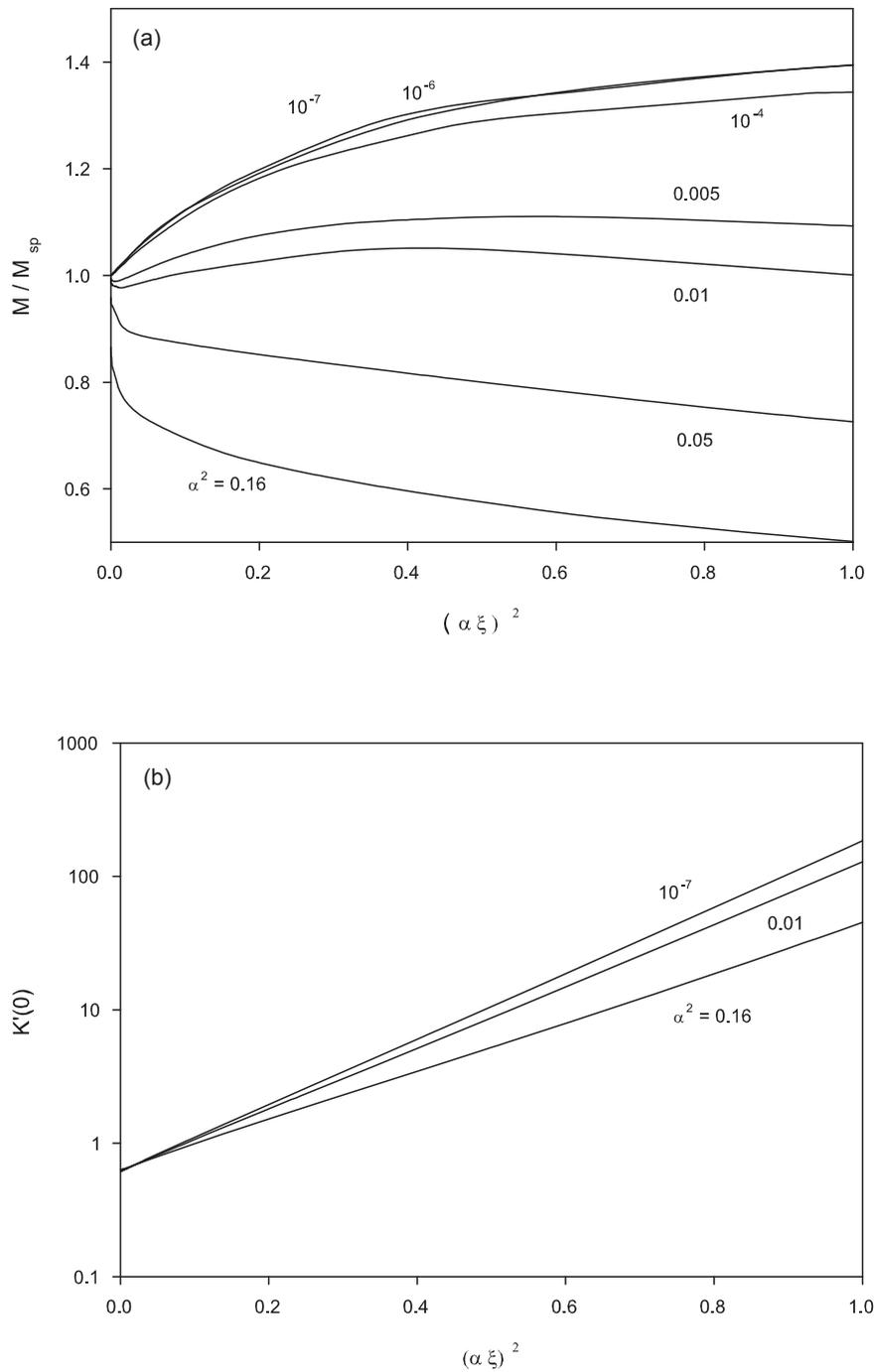}}
%%%\subfigure[Particle number, Mass and mean radius]
%%%{\label{hyper_r}\includegraphics[width=8cm]{hyper.eps}}
\end{center}
\caption{The sphaleron mass and the central Higgs field slope $K'(0)$ as a function  $(\alpha\xi)^2$ for $\beta=0.5$ and several values of $\alpha$.
\label{SphalMass}
}
\end{figure}

The solutions seem to obey the following scenario which is summarized by
Fig.\ref{SphalMass}~:  the relevant parameter is  $\tau \equiv (\alpha \xi)^2$.
As long as $\tau < 0.1$ the solutions deviate only very little from the ``probe limit''
i.e. from the case $\alpha = 0$ where gravity and the YMH system decouple. For $\tau  >  0.1$
we observe a very quick increase of the value $K'(0)$, as demonstrated by the figure and, as a consequence, of the value $R(0)$.   At the same time, there is a significant increase of the mass of the sphaleron.
Our  numerical solution is reliable up to  $K'(0) \sim 500$,
 this is the reason the lines in the figure somehow stop but this limitation is not harmful,
%%
%%%We cannot obtain significant numerical result
For the   value of the parameter $\alpha$ corresponding to the electroweak scale
$\alpha \simeq 10^{-17}$ it is very likely that  the phenomenological value $\xi \sim - 10^4$ will still
correspond to the 'probe-limit' regime and that the energy barrier will be of the same order of magnitude as the
corresponding sphaleron in flat space, i.e. $M_{sphaleron} \sim 3.63 M_W/ \alpha_w$. So the non-minimal coupling of the Higgs field to gravity should not affect the fermion number non-conserving exchange expected at the electroweak phase transition.

\clearpage

\subsection{Excited solutions}
 For completeness we  also discuss the family corresponding to $k=2$, i.e. 2 zeroes of $f(r)$.

%%Next, we study the response of the $k=2$ family of solutions to the non-minimal coupling.
%%The boundary conditions charaterizing excited (i.e. $k=2$) solution  from the sphaleron branch
%%was mentionned above. In the minimal case (and for $\beta$ fixed) the excited solutions form
%%two branches existing for $\alpha \in ]0, \alpha_c(2)[$

\begin{figure}[h]
\begin{center}
%%%\subfigure[Profile]
{\includegraphics[width=12cm]{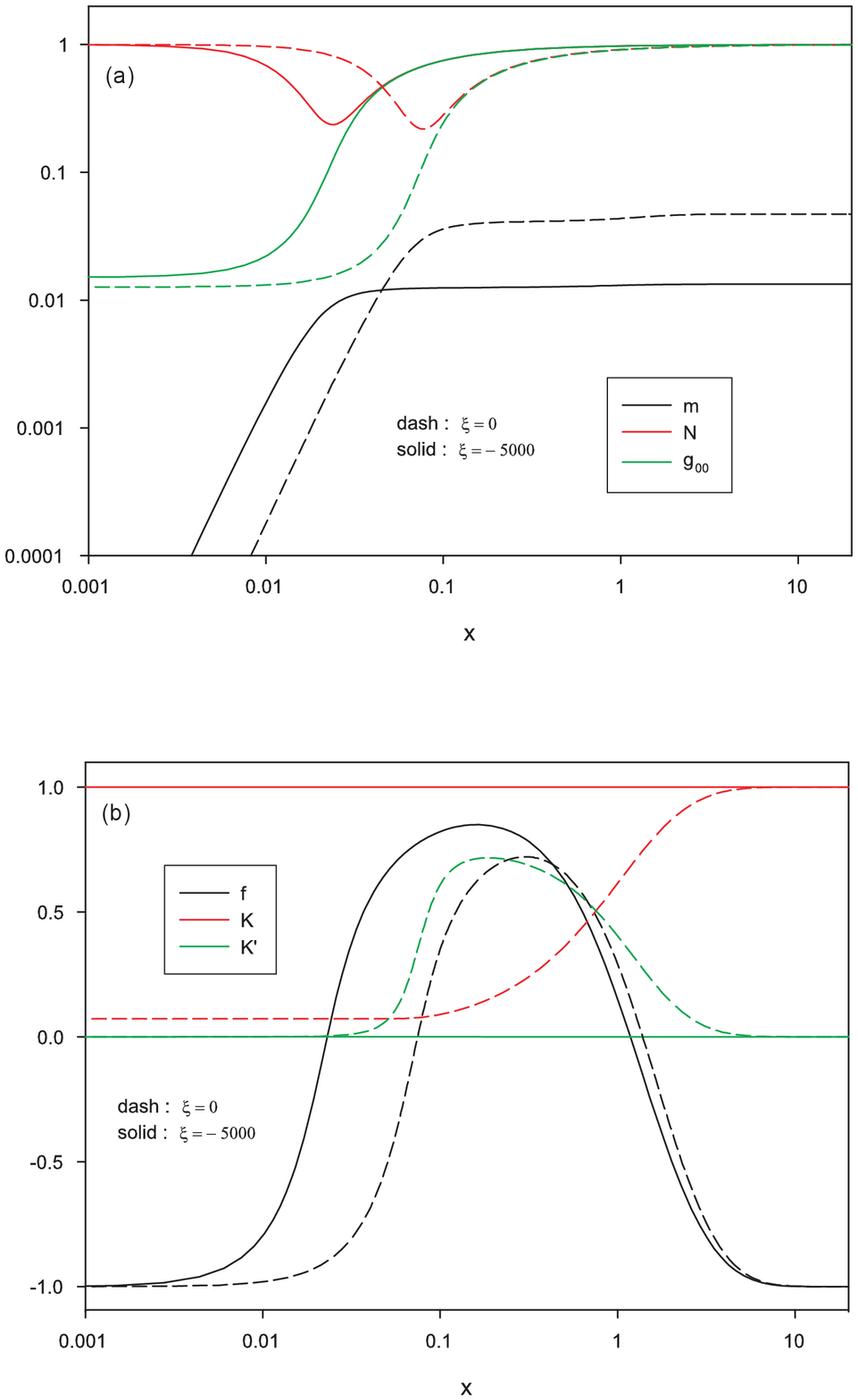}}
%%%\subfigure[Particle number, Mass and mean radius]
%%%{\label{hyper_r}\includegraphics[width=8cm]{hyper.eps}}
\end{center}
\caption{Profiles of a typical $k=2$ sphaleron for minimal ($\xi = 0$) and non-minimal coupling with $\xi = - 5000$. In both cases $\alpha = 0.05$ and $\beta=0.5$. (a) metric functions; (b) YMH fields.
 Note the logarithmic $x$-scale.
\label{inflation_1}
}
\end{figure}

%%we limited the analysis to the branch of solutions with the lowest energy.
In the presence of the non-minimal coupling and in particular for decreasing negative $\xi$,
it turns out that  the scalar field $K(r)$ approaches {\it uniformly}
the constant function $K(r)=1$ for $r\in [0, \infty]$. By contrast,
 the gauge field $f(r)$ keeps its own structure as  illustrated in Fig. \ref{inflation_1}
 for $\xi = 0$ and $\xi = - 5000$ (here the value $\alpha = 0.05$ is set since these solution exist for small $\alpha$ only).

\begin{figure}[h]
\begin{center}
%%%\subfigure[Profile]
{\includegraphics[width=12cm]{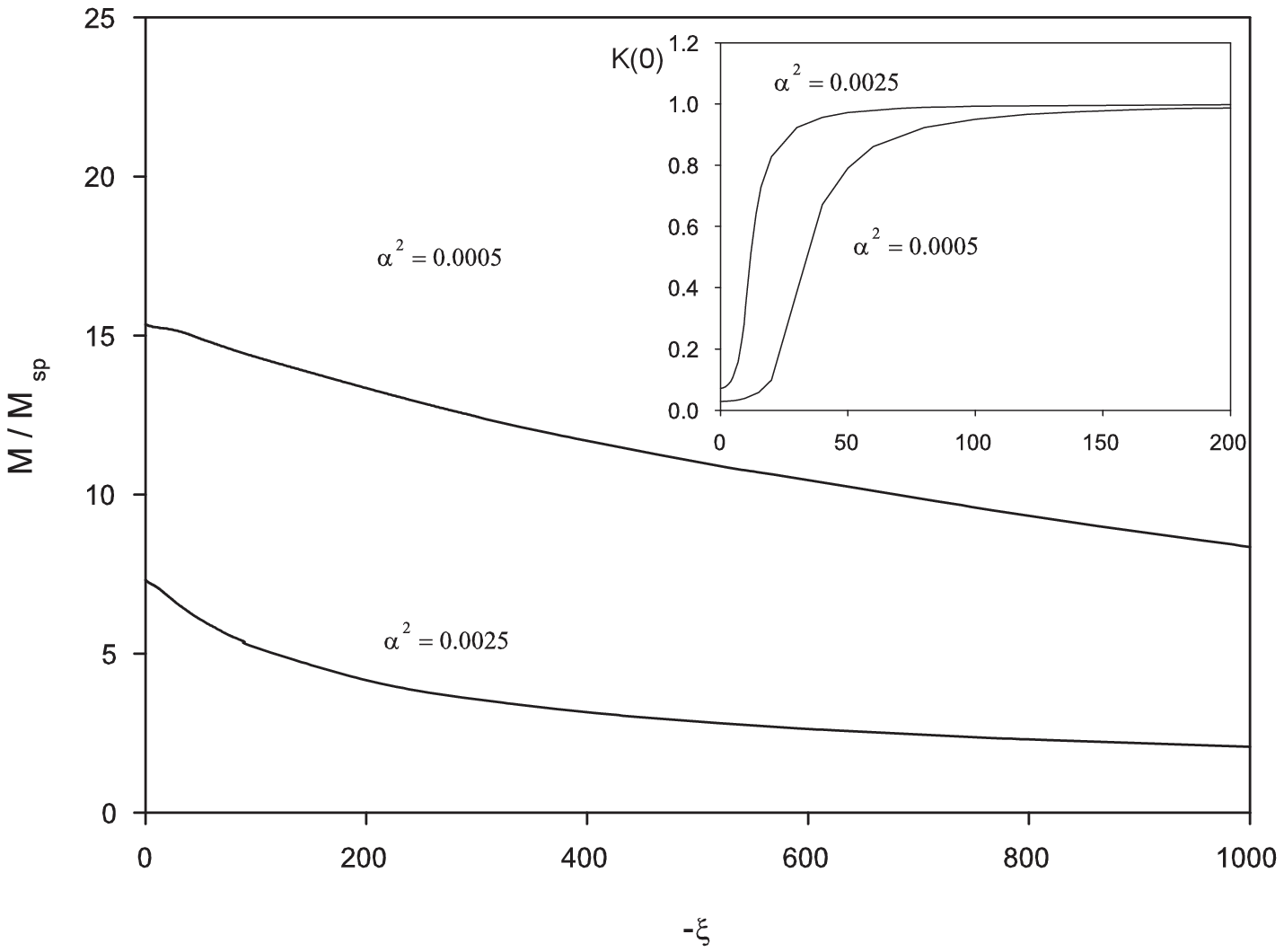}}
%%%\subfigure[Particle number, Mass and mean radius]
%%%{\label{hyper_r}\includegraphics[width=8cm]{hyper.eps}}
\end{center}
\caption{The dependence of the mass and of the value $K(0)$ on $\xi$ for two values of $\alpha$. Small values of $\alpha$ are chosen since these solution exist for small $\alpha$ only - see text. $\beta=0.5$.
\label{data_excited}
}
\end{figure}

This demonstrates that the solutions of the $k=2$ family, once strongly non-minimally coupled to gravity,
approach a kind of Yang-Mills-gravity configuration. Interestingly, this solution is not exactly a member
of the Bartnik-MacKinnon series because of the term $K^2(f+1)^2$ in the Lagrangian. The limiting solution
 rather describes a `massive' Yang-Mills-gravity system.
It should be stressed that the boundary conditions fulfilled by the $k=2$ solutions are the same
as the ones of the vacuum solution. One consequence of this is that, for $k=2$,
robust numerical  solutions  can be constructed
for much higher values of $-\xi$ than for the case $k=1$.
We have solved the system for several values of $\alpha$ and found some features
illustrated by Fig.\ref{data_excited}  for  $\alpha^2=0.0025$ and $\alpha^2 = 0.0005$~:
(i) the value $K(0)$ approaches slowly to  the value $K(0)=1$ when $\xi$ decreases; (ii)
 at the same time the mass deceases monotonically. As far as we could see, however
  the mass of the $k=2$ solutions  always remains higher than the energy barrier
  determined by the corresponding sphaleron (i.e. for $k=1$).

In summary, our numerical results strongly suggest that, when the non minimal coupling of the Higgs
field to gravity  becomes very strong, the solutions of the Einstein-Yang-Mills-Higgs equations
approach configurations where the Higgs field deviates only a little from its expectation value
$|\Phi| = v$ while the Yang-Mills field keeps a non trivial structure similar  to the Bartnik-McKinnon series. In the case $k=1$, the Higgs field is forced to be null at the origin because of the boundary conditions and the function $K(r)$ approaches a step function.
In the case $k=2$ the Higgs field approaches its expectation value uniformly on  space-time.

%%%%\subsection{Sphaleron inside a boson star}
%%%%Here we consider a boson star with a scalar field (minimally coupled) to the gravity through
%%%% the solutions discussed in the above section.

\clearpage

\section{The Magnetic Monopole}
\label{Numerical-Monopole}
\setcounter{equation}{0}
Next to be studied is the coupled system corresponding to a triplet of scalar fields (Georgi-Glashow SU(2) model) producing the magnetic monopole.

Substituting the above ansatz - Eq (\ref{Sphaleron+Monopole}) for isospin 1 into the field equations and performing some simplifications, yield the following system for the 4 unknown functions $A(r)$, $N(r)$, $H(r)$ and $f(r)$. For the YMH system we obtain now:
\begin{eqnarray}\label{MonHiggsEqs}
 \frac{(r^2AN H')'}{r^2 A}-\frac{2f^2}{r^2} H - \frac{\lambda}{4} v^2 (H^2-1)H +
 \, \hspace {6.0cm} \\ \nonumber
 \xi \frac{(6\xi-1)[N(H')^2+\frac{2f^2 H^2}{r^2}]+\frac{\lambda}{4} v^2 (H^2-1)[(6\xi-1)H^2+1]}{1/\kappa v^2 + \xi(6\xi-1)H^2}H=0
\end{eqnarray}
and
\begin{equation}
\frac{(ANf')'}{A}-(ev)^2 H^2 f - \frac{(f^2-1)f}{r^2}=0
\label{MonGaugeEqs}.
\end{equation}
while the (00) and (11) gravitational field equations are:
\begin{eqnarray}
\frac{N'}{r}+\frac{N-1}{r^2}+\frac{\kappa\left(\varepsilon_0+\varepsilon_1+u_0 +u_1+w_2-\xi(\tau_0+4\varepsilon_0-\tau_1)\right)}{1-\xi\kappa v^2 H^2}=0
\label{MonEinsEq00}
\end{eqnarray}
\begin{eqnarray}
\frac{N'}{r}+\frac{N-1}{r^2}+\frac{2N}{r}\frac{A'}{A}+\frac{\kappa\left(-\varepsilon_0-\varepsilon_1+u_0 +u_1+w_2-\xi\tau_2\right)}{1-\xi\kappa v^2 H^2}=0
\label{MonEinsEq11}
\end{eqnarray}
where now
\begin{eqnarray}
\varepsilon_0=\frac{v^2N(H')^2}{2} \,\,\,\,\,\,&;& \,\,\, \varepsilon_1=\frac{N(f')^2}{e^2 r^2} \\ \nonumber
u_0=\frac{\lambda v^4}{16} (H^2-1)^2  \,\,\,\,\,\,&;& \,\,\, u_1=\frac{(f^2-1)^2}{2e^2 r^4} \,\,\,\,\,\,; \,\,\, w_2=\frac{v^2 f^2 H^2}{r^2}\\ \nonumber
\tau_0=\frac{2v^2 H(r^2AN H')'}{r^2 A} \,\,\,\,\,\,&;& \,\,\,\tau_1=2v^2NHH'\left(\frac{A'}{A}+\frac{N'}{2N}\right)
\,\,\,\,\,\,; \,\,\,\tau_2=\tau_1+\frac{4v^2 NHH'}{r}
\end{eqnarray}
Note the differences with respect to the sphaleron case in some numerical factors and the replacements $K\rightarrow H$ and $(f+1)\rightarrow 2f$. We also use $w_2$ instead of the analogous $u_2$.

As for the sphaleron, the monopole mass $M$ is extracted from the asymptotic behavior of $N(r)$: $GM={\cal M}(\infty)$ or from the analogous integral expression:
\begin{eqnarray}
M=4\pi \int_0^{\infty} dr \hspace {0.15cm} r^2 \hspace {0.1cm} \frac{\varepsilon_0+\varepsilon_1+u_0 +u_1+w_2-\xi(\tau_0+4\varepsilon_0-\tau_1)}{1-\xi\kappa v^2 H^2}
\label{MassMonopols}
\end{eqnarray}

In dimensionless form we have the following
\begin{eqnarray}\label{MonHiggsEqsFDless}
 \frac{(x^2AN H')'}{x^2 A}-\frac{2f^2}{x^2} H - \frac{\beta}{2} (H^2-1)H +
 \, \hspace {6.0cm} \\ \nonumber
 4\xi \alpha^2 \frac{(6\xi-1)[N(H')^2+\frac{2f^2 H^2}{x^2}]+\frac{\beta}{2} (H^2-1)[(6\xi-1)H^2+1]}{1 + 4\xi \alpha^2 (6\xi-1) H^2}H=0
\end{eqnarray}
\begin{equation}
\frac{(ANf')'}{A}- 2H^2 f - \frac{(f^2-1)f}{x^2}=0
\label{MonGaugeEqsDless}.
\end{equation}
The (00) and (11) gravitational field equations are:
\begin{eqnarray}
\frac{N'}{x}+\frac{N-1}{x^2}+\frac{2\alpha^2}{1-4\xi\alpha^2 H^2}\left[N(H')^2+\frac{N(f')^2}{x^2}+\frac{\beta}{4} (H^2-1)^2 +\frac{(f^2-1)^2}{2x^4} \nonumber \right. \\ \left.  + \frac{2f^2 H^2}{ x^2}-4\xi \left(\frac{H(x^2AN H')'}{x^2 A}+N(H')^2-NHH'\left(\frac{A'}{A}+\frac{N'}{2N}\right)\right)\right]=0
\label{MonEinsEq00Dless}
\end{eqnarray}
\begin{eqnarray}
\frac{N'}{x}+\frac{N-1}{x^2}+\frac{2N}{x}\frac{A'}{A}+\frac{2\alpha^2}{1-4\xi\alpha^2 H^2}\left[-N(H')^2-\frac{N(f')^2}{x^2}+\frac{\beta}{4} (H^2-1)^2 \nonumber \right. \\ \left. +\frac{(f^2-1)^2}{2x^4} +\frac{2f^2 H^2}{x^2}-4\xi NHH' \left(\frac{A'}{A}+\frac{N'}{2N}+\frac{2}{x}\right)\right]=0
\label{MonEinsEq11Dless}
\end{eqnarray}

 As before, $H''$ can be converted to first order by using the field equation (\ref{MonHiggsEqsFDless}) for $H$ and writing:
\begin{eqnarray}
\frac{H(x^2AN H')'}{x^2 A}= \frac{\frac{2 f^2}{x^2}-\frac{\beta}{2} \left( 4\xi \alpha ^2 -1\right) \left(H^2-1\right)  -  4\xi \alpha ^2 (6 \xi -1) N
   \left(H'\right)^2}{1+  4\xi \alpha ^2 (6 \xi -1) H^2 } H^2
\label{H''substit}
\end{eqnarray}

\begin{figure}[h]
\begin{center}
%%%\subfigure[Profile]
{\includegraphics[width=12cm]{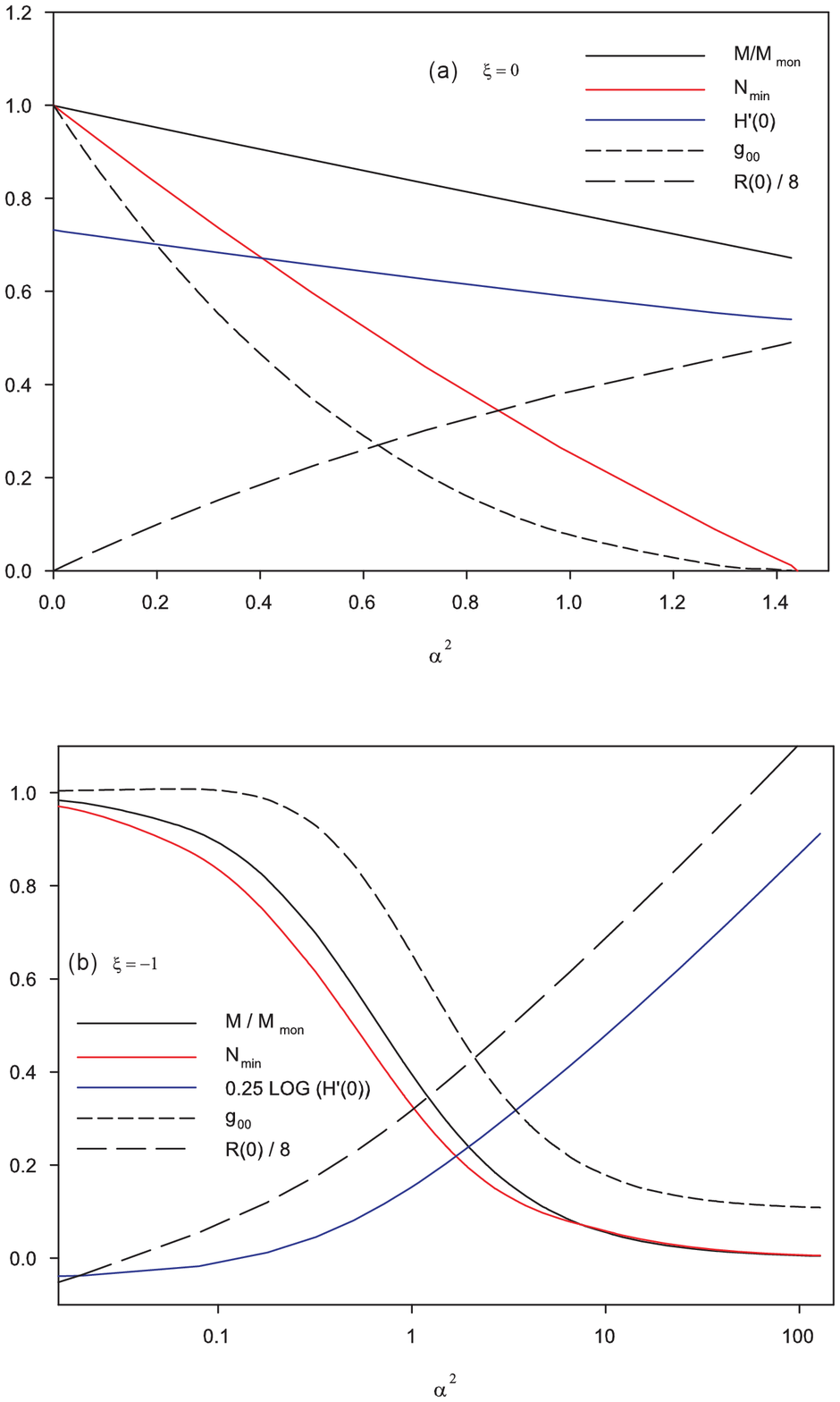}}
%%%\subfigure[Particle number, Mass and mean radius]
%%%{\label{hyper_r}\includegraphics[width=8cm]{hyper.eps}}
\end{center}
\caption{Several quantities characterizing the monopole as functions of the parameter $\alpha^2$ for $\beta=0.5$ and: (a) $\xi = 0$; (b)  $\xi = - 1$ .
\label{data_mono}
}
\end{figure}

In the case of a minimal coupling ($\xi=0$), the following scenario was  discovered by Breitenlhoner, Forgacs and Maison (BFM) \cite{BreitenlohnerEtAl1991}: The flat magnetic monopole is progressively deformed by gravity, the function $N(r)$ develops a local
minimum which gets deeper while $\alpha$ increases. At a maximal value of $\alpha$ (say $\alpha = \alpha_{cr}(\beta)$),
we get $N(r_h)=0$ for some finite value of $r_h$ and the gravitating monopole bifurcates into an extremal
Reisner-Nordstrom black hole with $r_h$ as an horizon. This scenario seems to persist for $\xi >0$ but in this paper
we set the main emphasis on the case $\xi < 0$ and more especially for $-\xi \sim 10^4$.
Let us first point out that for $\xi < 0$ (at least for $\xi \leq -1$), the BFM scenario does not seem to hold.
Instead, the {\it non-minimally}-gravitating monopole seems to exist for large values of $\alpha$. Increasing $\alpha$
the numerical integrations of the equations reveals that the value $N_m$ slowly goes to zero while the value $H'(0)$
slowly increases. This is illustrated by Fig.\ref{data_mono} for $\xi = 0, -1$.
It is not clear whether a critical phenomenon  stops the solution to exist at a finite value
of $\alpha$. We suspect that the solution exist for arbitrary values of $\alpha$.

A general feature of the solutions seems to be that, for fixed $\alpha$ and decreasing $\xi$,
the parameter $H'(0)$ increases considerably so that
 the function $H(r)$ varies considerably around the origin and quickly reaches its asymptotic value $H=1$.
As a consequence, the numerical construction of the solutions
becomes  tricky for large values of $-\xi$.

\begin{figure}[h]
\begin{center}
%%%\subfigure[Profile]
{\includegraphics[width=12cm]{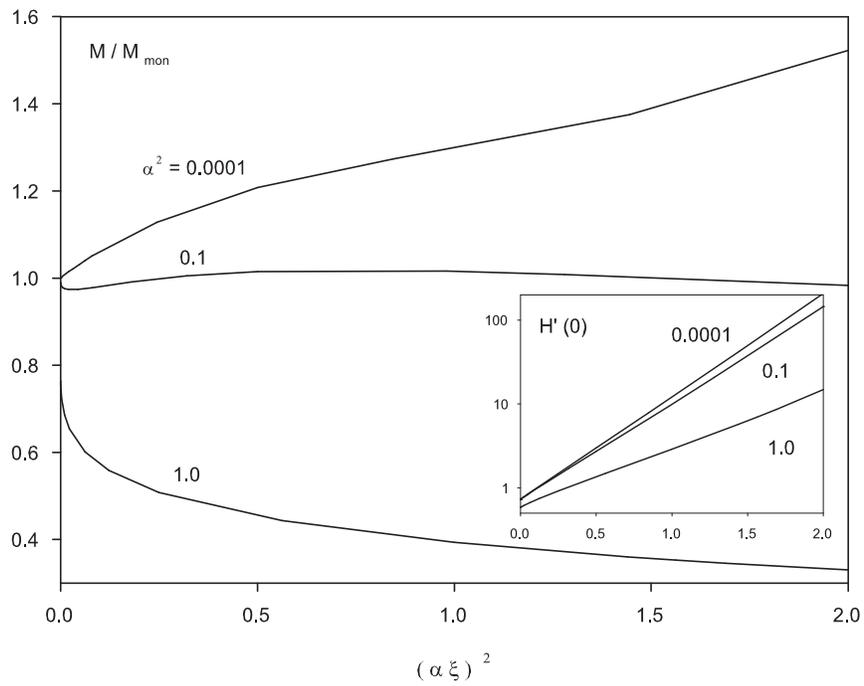}}
%%%\subfigure[Particle number, Mass and mean radius]
%%%{\label{hyper_r}\includegraphics[width=8cm]{hyper.eps}}
\end{center}
\caption{ Several quantities characterizing the monopole as functions of
 $(\alpha\xi)^2$  for $\alpha = 0.0001$ and $\beta = 0.5$.
\label{large_xxi}
}
\end{figure}

\begin{figure}[h]
\begin{center}
%%%\subfigure[Profile]
%\label{inflation}
{\includegraphics[width=12cm]{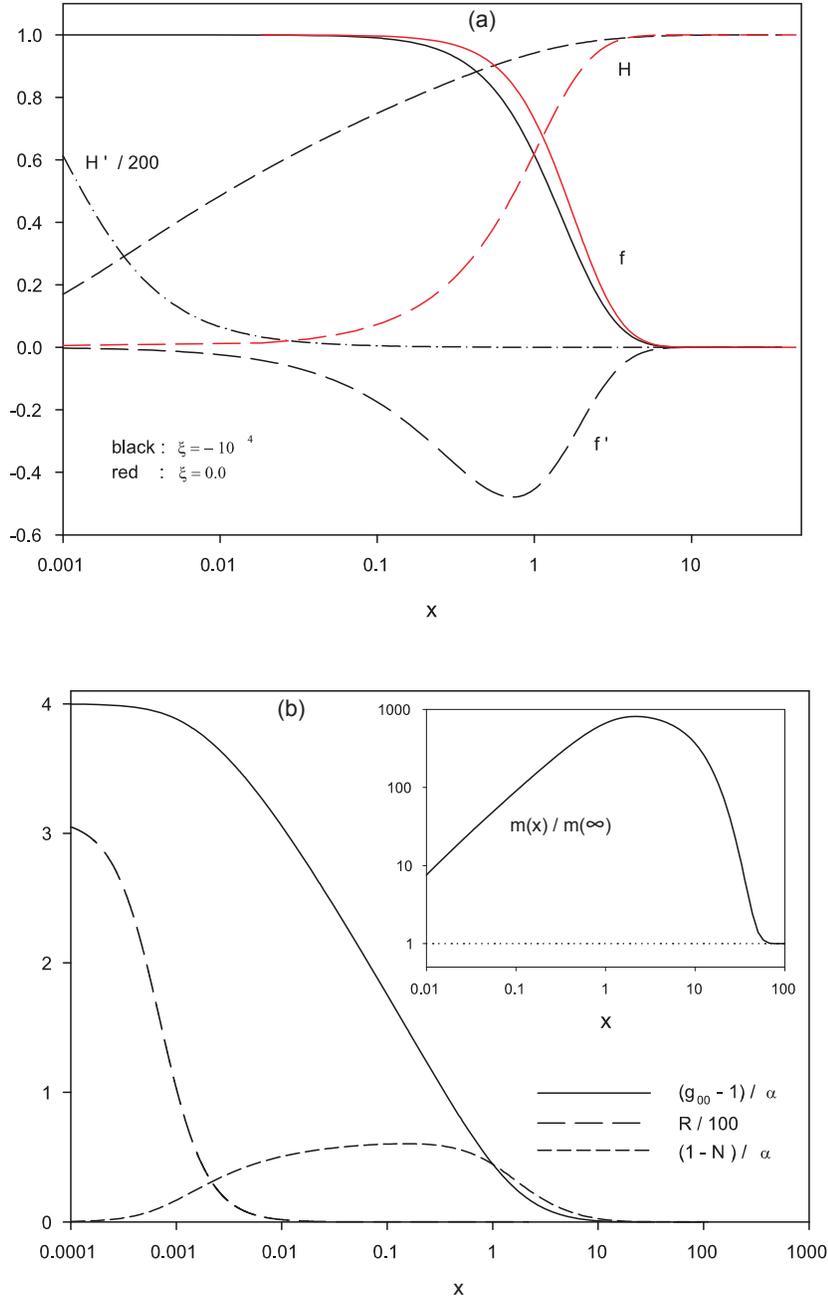}}
%%%\subfigure[Particle number, Mass and mean radius]
%%%{\label{hyper_r}\includegraphics[width=8cm]{hyper.eps}}
\end{center}
\caption{(a) Profiles of the flat  Monopole (red curves) and of the non minimally coupled monopole (black curves) with $\xi = - 10^4$. In both cases $\alpha = 0.0001$ and $\beta = 0.5$. (b) Deviation of the metric functions $N,A$ with respect to
Minkowski space and the Ricci scalar around the origin.
\label{inflation}
}
\end{figure}

We also found that the larger the value of $\alpha$ is, the quicker the numerical problems occur.
Fortunately, for small values of $\alpha$ (i.e. in the physically realistic domain),
solutions can be constructed up to $\xi \sim - 10^4$  without any numerical problem.
Our Fig.\ref{large_xxi} illustrates the dependance of the mass and the central Higgs slope on $\xi$ (here $\alpha = 0.0001$
was set). The profiles of the monopole fields for non-minimal coupling  is displayed in Fig.\ref{inflation}.
In order to appreciate the influence of the non minimal coupling on the matter fields,
the profiles of $f, H$ corresponding to the minimally coupled solution are superposed in the upper part of the figure.
The metric function $N,A$
hardly differ from the Minkowski space-time but their second derivative differs strongly, leading to a very significant deviation of the curvature with respect to $R=0$. In Fig.\ref{inflation} we supplement the derivatives $f',H'$ for the $\xi = -1000$ case to demonstrate  the strong variation of the scalar function around the center.
The deviation from Minkowski is of order $\alpha$ and is also illustrated by Fig.\ref{inflation}.

%%%%
%%%%
%%%%
%%%
%%%
%%%
\clearpage
\section{Sphalerons Inside Boson Stars }
\label{S-BS}
In the previous sections, we  studied the effects of the non-minimal coupling of the Higgs field
on the sphaleron and monopole solutions and constructed the associated space-times.
It is also natural to extend the analysis to compact star-like objects that interact gravitationally with these solutions. In this respect, boson stars and Q-stars (see e.g. \cite{Lynn1988} for a review) made out of an (ungauged)
complex scalar field minimally coupled to gravity but self interacting are perhaps the simplest examples.
With this motivation, we considered  the field equations of the Lagrangian density (\ref{TotLag}) extended by a boson star sector
\be
     {\cal L}_{add} ={\cal L}_{BS} = \partial_{\mu}\phi^*  \partial^{\mu}\phi - U(|\phi|)
\ee
In many papers reporting   Q-balls \cite{Volkov:2002aj} and
minimally coupled boson stars (see e.g. \cite{Kleihaus:2005me},\cite{Kleihaus:2011sx}),
the self interacting potential is chosen to be
\be
       U(|\phi|) = c_6 |\phi|^6 - c_4 |\phi|^4 + m_B^2 |\phi|^2
\ee
where $m_{B}$ represents the mass of the new boson.  The choice $c_6=1/v^2$, $c_4 = 2$, $m_B^2/v^2 = 1.1$ is often used in the literature and we adopt the same potential in order to calibrate our solutions with  existing ones.
Apart from their mass, the boson stars are characterized by the conserved global charge $Q$ associated with the phase-change symmetry of the scalar field and the resulting conserved current. The charge  is computed through the following integral
\be
    Q = \int d^3 x  |g|^{1/2} j^0  \ , \ \
    j^{\mu} = -i(\phi^* \partial^{\mu} \phi - \phi \partial^{\mu} \phi^* ) \  .
\ee
 The charge $Q$ is interpreted as the particle number of the boson star;  a binding
  energy can then  be defined for the boson star as the difference between the mass of $Q$ elementary bosons of mass $m_B$, and the mass of the boson star. For the mass of the boson star we take the difference $M_T-M_{sp}(k)$  where $M_T$ is the total gravitational mass extracted from the decay of the metric field $N(r)$ and $M_{sp}(k)$ is the mass of the self-gravitating sphaleron-like solutions calculated in sec. \ref{Numerical-Sphaleron}. The binding energy is thus   defined according to $B= m_B Q - M_T + M_{sp}(k)$. Configurations with $B > 0$ are expected to be stable.
  \par
  Completing the spherically symmetric ansatz used above by
 \be
       \phi = v F(r) e^{i \omega t}
 \ee
 the field equations of the full system
 lead to an extra differential equation for the function $F(r)$. This equation has to be solved
 with the boundary conditions $F'(0)=0$ (necessary for the regularity at the origin)
 and $F(\infty)=0$ (for a finite mass and charge).
 Only for a finite interval of values of the frequency $\omega$ do such solutions exist. Although there
 is a relation  between the frequency $\omega$ and the central value $F(0)$,
  this relation is not one to one~: two or more solutions can exist with the same frequency but different
 values of $F(0)$. With the above ansatz,  the charge $Q$ is computed though the following integral
\be
    Q = 8\pi \omega v^2 \int_{0}^{\infty} dr \hspace {0.15cm} r^2  F^{2}(r)/A(r)N(r) .
\ee
 In the following, we discuss the
 boson stars coupled to the solutions associated with the Higgs doublet of sections \ref{FEqs-Sphaleron}-\ref{Numerical-Sphaleron}
   (i.e. the vacuum $k=0$, the sphaleron $k=1$  and the excited solution $k=2$).
\par
 Considering first the vacuum solution of the YMH sector $K(r)=1, f(r)=-1$,
%%% (it is natural to characterize this solution by $k=0$)
 the equation for the boson star
 reduces to standard equation with an effective coupling $1/\kappa  \rightarrow  1/\kappa + |\xi| v^2$.
  In \cite{Kleihaus:2005me}, it was shown in particular that boson stars have a non trivial limit in the
 infinite limit of the parameter $\kappa$. So  solutions relevant for the Higgs inflation
 do exist, their properties can be found in \cite{Kleihaus:2005me}.

 Once coupled to the $k=1$ and $k=2$ sphalerons, the solutions, to our knowledge, have not been discussed before.
 Examining first the case $\xi = 0$, we constructed  families of boson stars coupled to the sphaleron
  and its excited version.  They both exist in an interval of $\omega \in [\omega_m, \omega_M]$.
  The  solutions on the various branches can be characterized by the frequency $\omega$ or, alternatively,
  by the central value of the scalar field $F(0)$. The limit $F(0) \to 0$ corresponds to the limit $\omega \to \omega_M$.
  The function $F(r)$  approaches the null function uniformly in this limit.
    The maximal value
  $\omega_M$ is roughly insensitive to the presence of the YMH fields.
  (Note: the behavior of $\omega_m$ will be discussed in a future report).

\begin{figure}[t]
\begin{center}
%%%\subfigure[Profile]
%\label{inflation}
{\includegraphics[width=12cm]{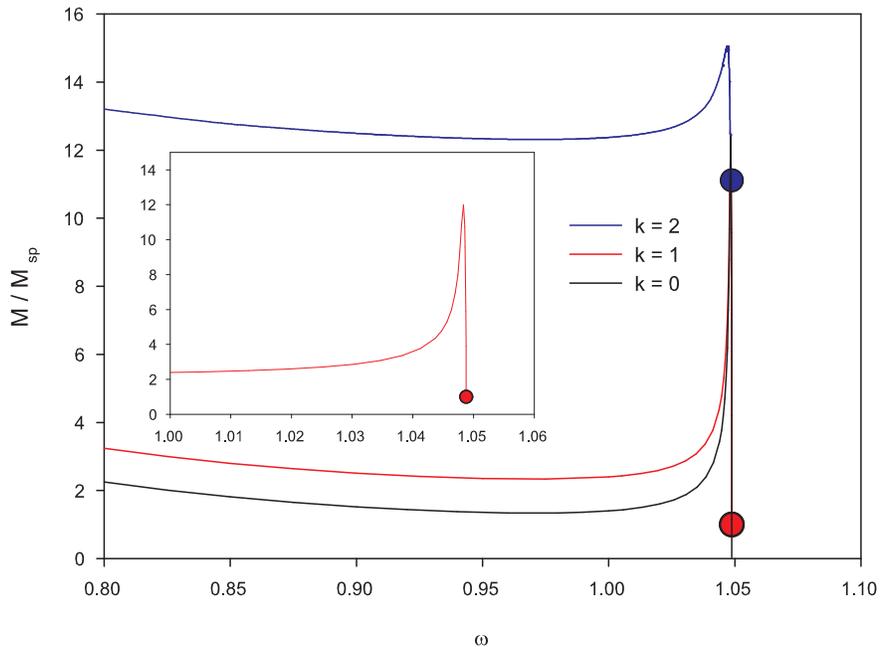}}
%%%\subfigure[Particle number, Mass and mean radius]
%%%{\label{hyper_r}\includegraphics[width=8cm]{hyper.eps}}
\end{center}
\caption{Mass dependence on $\omega$ (in units of $ev/\sqrt{2}$) for the sphaleron-boson star corresponding to $k=0,1,2$ for $\alpha^2 = 0.001$, $\beta=0.5$. The bullets indicate the corresponding sphalerons.
\label{data_boson_star_w_M}
}
\end{figure}

  The mass dependance of the sphaleron-boson star on the frequency $\omega$ is shown in Fig.\ref{data_boson_star_w_M}. The horizontal axis gives $\omega$ in units of $ev/\sqrt{2}$, i.e.
  $\bar{\omega}=\sqrt{2}\omega/ev$.
  As expected, the masses of the solutions corresponding to $k=0,1,2$ are increasing with $k$. The analysis
  of this plot shows that on each branch,   the mass present a local maximum just
  before reaching the value of the corresponding sphaleron for $\omega = \omega_M \approx 1.05 ev/\sqrt{2}$. The bullets represent
  the masses of the $k=1,2$ sphalerons.
  The black line reaches the local maximum at $M \approx 12 M_{sp}$ and, of course, stops at $M=0$.
%%%

 \begin{figure}[t]
\begin{center}
%%%\subfigure[Profile]
%\label{inflation}
{\includegraphics[width=12cm]{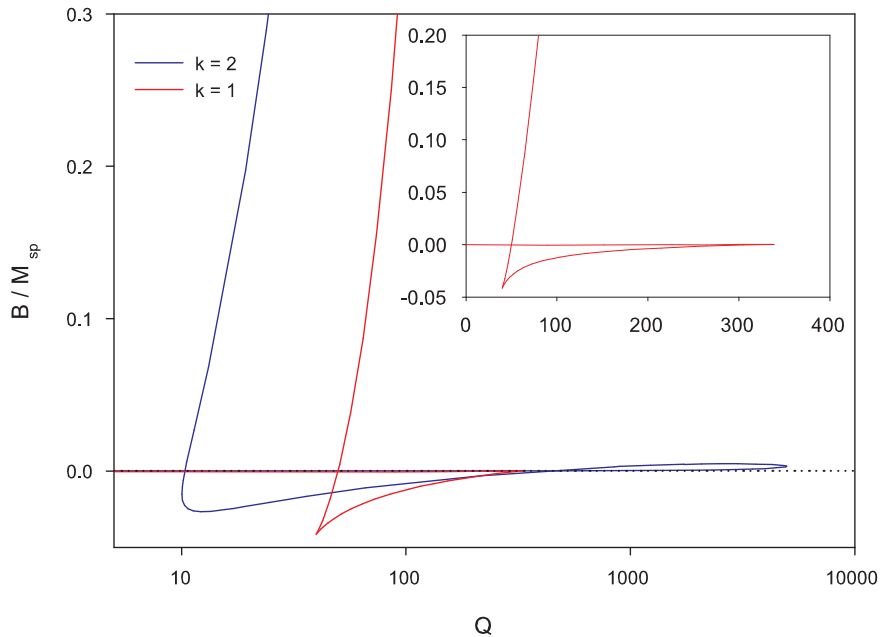}}
%%%\subfigure[Particle number, Mass and mean radius]
%%%{\label{hyper_r}\includegraphics[width=8cm]{hyper.eps}}
\end{center}
\caption{Binding energy of the $k=0,1$ (virtually indistinguishable) and $k=2$ solutions as function of $Q$  for $\alpha^2 = 0.001$,  $\beta=0.5$.
The insert shows the details of the $k=1$ curve in the region of small $Q$.
\label{data_boson_star_Q_B}
}
\end{figure}

 \begin{figure}[t!]
\begin{center}
%%%\subfigure[Profile]
%\label{inflation}
{\includegraphics[width=12cm]{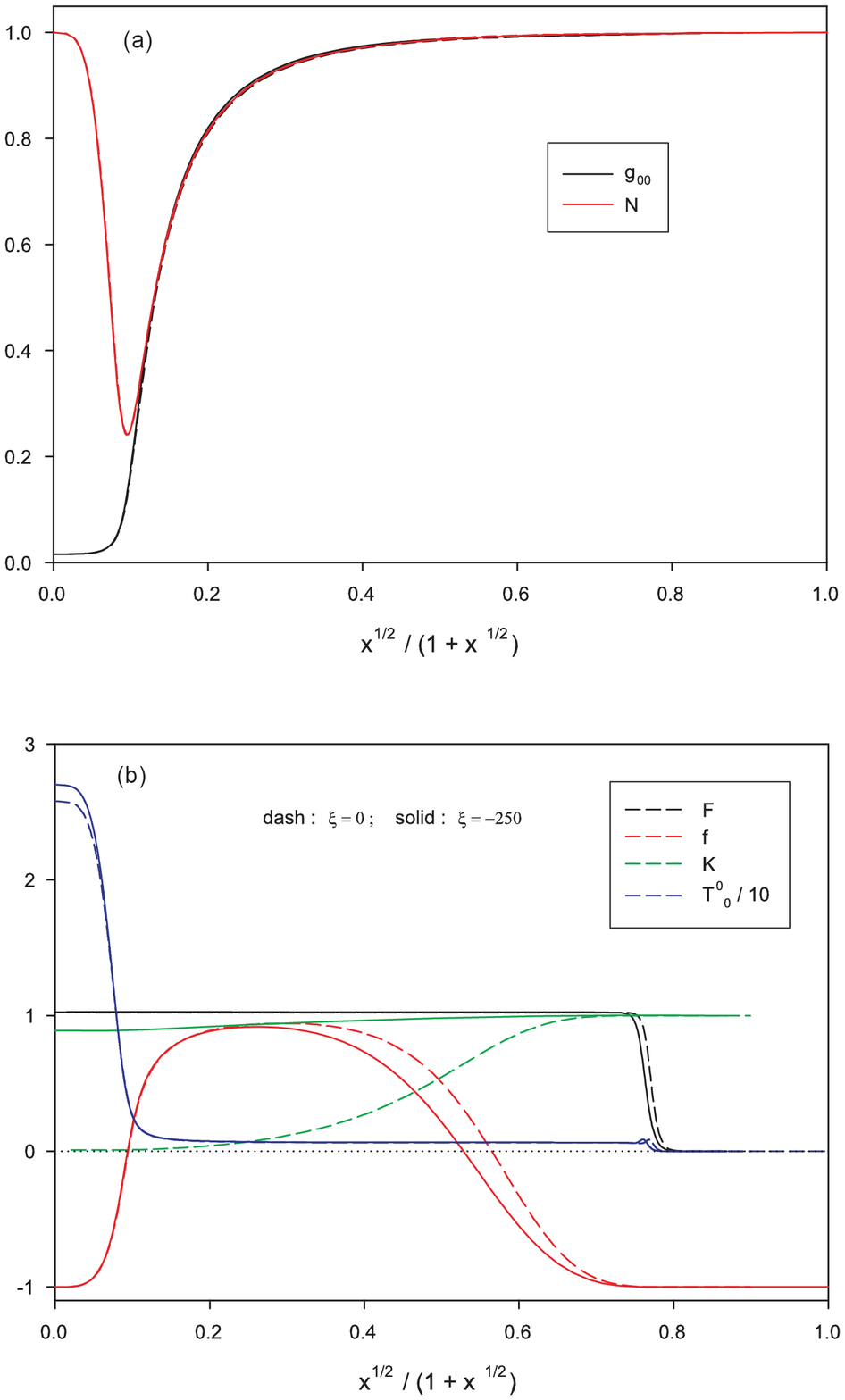}}
%%%\subfigure[Particle number, Mass and mean radius]
%%%{\label{hyper_r}\includegraphics[width=8cm]{hyper.eps}}
\end{center}
\caption{Comparison of the profiles of a stable $k=2$ sphaleron-boson star
corresponding to  $\alpha = 0.0001$, $\beta=0.5$,  $\bar{\omega} = 0.45$ and $\xi = 0; - 250$.
(a) metric components ; (b) matter functions. The difference in the metric components is hardly visible.
\label{profile_k_2_bs_mn}
}
\end{figure}

The question which raises naturally is the question of stability of these solutions.
In the flat limit, the sphaleron presents one unstable mode.
The $k$-node solution of the Bartnik-McKinnon family present $k$ unstable modes \cite{GaltsovVolkov}
and it can be expected that, on the lowest
branches of Fig.\ref{comparaison_k_1_2} the $k=1,2$ gravitating sphalerons
 characterized by $k=1$ and $k=2$ present $k$ unstable modes.

 The simplest way to address the question
 of stability is to examine the value of the binding energy defined above as  $B= m_B Q - M_T + M_{sp}(k)$. In correspondence to the
 three branches of solutions presented in Fig.\ref{data_boson_star_w_M}, the binding energy is shown  in Fig.\ref{data_boson_star_Q_B}.
% It turns out that the binding energy of the solutions becomes positive
% (suggesting stablity) for large enough values of $Q$ (and therefore of the Mass)
% suggesting that a sufficiently large mass is necessary to stabilize the sphaleron-boson lumps.
Along the case of pure boson star the Q-B plot case of sphaleron boson star reveals a succession of connected  branches forming in cusp-catastrope curve at their intersection (in fact both cases are hardly distinguished on the plot). The corresponding plot in the case of $k=2$ solution  reveals several branches presenting the same qualitative pattern; however the curve seems to be smooth.
 In terms of the frequency, the stability
 occurs when the parameter $\omega$ is smaller than a critical value, $\omega_c$. For the case in the figure, we find
 $\bar{\omega}_c = 0.9$, $\bar{\omega}_c = 0.88$ and
   $\bar{\omega}_c = 0.85$ respectively for $k=0,1,2$.  For the small values of $\omega$,
   the scalar field of the boson star is in the so called thin wall limit: the profile of the function $F(r)$
   is constant on a large interval starting at $r=0$ and  decreases sharply to zero around
   some radius, say $r = r_w$, forming a wall. Accordingly, the corresponding field $\phi$
   is concentrated in a sphere, forming a compact object.

Finally, we studied the response of the  boson stars to the non-minimally coupled sphaleron as well as monopole.  The problem in vast and  the details about these solutions will be presented elsewhere \cite{YBYV2015}.
However, we would like to sketch already here some results related to the $k=2$ sphaleron.
For this type of solution, the
 boundary conditions of the YMH fields
 are identical to those discussed in \cite{Fuzfa:2013yba,Schlogel:2014jea} where the gauge fields are not excited from the beginning.
 Our solution could therefore be
 interpreted as a gauged version of the lump constructed in \cite{Fuzfa:2013yba,Schlogel:2014jea} (the stictly
 compact star used \cite{Fuzfa:2013yba,Schlogel:2014jea}  is here mimicked by a boson star in the thin wall limit).
 Setting for definiteness
 $\alpha = 0.0001$, we were able to construct the solutions for large values of $-\xi$. It turns out that,
 decreasing $\xi$, the Higgs field tends to approach its expectation value in space-time while the
 gauge field keeps a non trivial sructure as shown in Fig. \ref{profile_k_2_bs_mn}.
 The boson-star field $|\phi|$ is
 practically constant in the sphere $x < x_w$ (in the case of the figure $x_w \sim 10$) and practically zero outside the sphere.
 The contribution of the boson star field to the energy density, $T^0_{0}$, is also presented; note
 the jump around $x_w \sim 10$.
  On the figure, we intentionally set
 $\xi = - 250$ in order to still have a small -but significant- deviation of the function $K$ from the constant value $K=1$. A more systematic analysis is currently in progress.

\section{Conclusion}
It is known for a long time that the sphaleron and the monopole,
two of the most studied classical solutions occuring in  Yang-Mills-Higgs field theory,
 persist when the Lagrangian is minimally coupled to gravity.
 The gravitating field equations even allow for extra solutions which do not have a flat space limit.
The recently proposed Higgs-inflation scenario raises the fascinating perspective
that one of the most intriguing element of particle physics, the Brout-Englert-Higgs boson,
could play the role of the inflaton through a non minimal coupling of the underlying scalar
field to gravity.

In this paper, we found convincing results showing that both, the monopole and the sphaleron
are also present as classical solutions when the Higgs field is non-minimally
coupled to gravity. If the Higgs-inflation scenario turns out to be viable, our results indicate
that the mass of the sphaleron in the relevant domain of the parameter space
       is not affected by the non minimal coupling. As such, the rate of baryon violating
      processes would not be affected by the new coupling.
Similarly, the mass of the monopole slightly changes too.
Although we checked it with a triplet of scalar fields of the Georgi-Glashow model, we
expect this feature to hold in more realistic GUT Lagrangians.

Extending the Lagrangian with an additional complex scalar field minimally coupled to gravity,
we manage to construct boson stars, mimicking star-like objets.
We argue that sphalerons can be trapped inside a boson star forming a sufficiently massive
system. These systems constitute localized objects and in a suitable domain of the
frequency, they are characterized by a positive binding energy.

The following  feature seems to be common for all the cases that we investigated~:
the decrease of the non minimal
coupling parameter $\xi$ forces the Higgs field to deviate only a little from its expectation value.
To the contrary, the Yang-Mills field keeps its structure.
 Likely, the classical lump constructed in  \cite{Fuzfa:2013yba} (where gauge
fields are not excited) cannot be approached with our parametrization of the fields of the theory.
Although both boson fields, $K(x)$ and $F(x)$, present similar properties and profiles, the
two solutions strongly differ from the gauge field.

\textbf{Acknowledgement}: The authors thank S. Schl\"{o}gel for helpful discussions.

\end{document}